\documentstyle[twoside,epsfig]{article}

\catcode`\@=11
\long\def\@makefntext#1{
\protect\noindent \hbox to 3.2pt {\hskip-.9pt  
$^{{\eightrm\@thefnmark}}$\hfil}#1\hfill}		%CAN BE USED 

\def\thefootnote{\fnsymbol{footnote}}
\def\@makefnmark{\hbox to 0pt{$^{\@thefnmark}$\hss}}	%ORIGINAL 
	
\def\ps@myheadings{\let\@mkboth\@gobbletwo
\def\@oddhead{\hbox{}
\rightmark\hfil\eightrm\thepage}   
\def\@oddfoot{}\def\@evenhead{\eightrm\thepage\hfil
\leftmark\hbox{}}\def\@evenfoot{}
\def\sectionmark##1{}\def\subsectionmark##1{}}

\oddsidemargin=\evensidemargin
\renewcommand{\thefootnote}{\fnsymbol{footnote}}

\newcounter{sectionc}\newcounter{subsectionc}\newcounter{subsubsectionc}
\renewcommand{\section}[1] {\vspace{12pt}\addtocounter{sectionc}{1} 
\setcounter{subsectionc}{0}\setcounter{subsubsectionc}{0}\noindent 
	{\tenbf\thesectionc. #1}\par\vspace{5pt}}
\renewcommand{\subsection}[1] {\vspace{12pt}\addtocounter{subsectionc}{1} 
	\setcounter{subsubsectionc}{0}\noindent 
	{\bf\thesectionc.\thesubsectionc. {\kern1pt \bfit #1}}\par\vspace{5pt}}
\renewcommand{\subsubsection}[1] {\vspace{12pt}\addtocounter{subsubsectionc}{1}
	\noindent{\tenrm\thesectionc.\thesubsectionc.\thesubsubsectionc.
	{\kern1pt \tenit #1}}\par\vspace{5pt}}
\newcommand{\nonumsection}[1] {\vspace{12pt}\noindent{\tenbf #1}
	\par\vspace{5pt}}

\newcounter{appendixc}
\newcounter{subappendixc}[appendixc]
\newcounter{subsubappendixc}[subappendixc]
\renewcommand{\thesubappendixc}{\Alph{appendixc}.\arabic{subappendixc}}
\renewcommand{\thesubsubappendixc}
	{\Alph{appendixc}.\arabic{subappendixc}.\arabic{subsubappendixc}}

\renewcommand{\appendix}[1] {\vspace{12pt}
        \refstepcounter{appendixc}
        \setcounter{figure}{0}
        \setcounter{table}{0}
        \setcounter{lemma}{0}
        \setcounter{theorem}{0}
        \setcounter{corollary}{0}
        \setcounter{definition}{0}
        \setcounter{equation}{0}
        \renewcommand{\thefigure}{\Alph{appendixc}.\arabic{figure}}
        \renewcommand{\thetable}{\Alph{appendixc}.\arabic{table}}
        \renewcommand{\theappendixc}{\Alph{appendixc}}
        \renewcommand{\thelemma}{\Alph{appendixc}.\arabic{lemma}}
        \renewcommand{\thetheorem}{\Alph{appendixc}.\arabic{theorem}}
        \renewcommand{\thedefinition}{\Alph{appendixc}.\arabic{definition}}
        \renewcommand{\thecorollary}{\Alph{appendixc}.\arabic{corollary}}
        \renewcommand{\theequation}{\Alph{appendixc}.\arabic{equation}}
        \noindent{\tenbf Appendix \theappendixc #1}\par\vspace{5pt}}
\newcommand{\subappendix}[1] {\vspace{12pt}
        \refstepcounter{subappendixc}
        \noindent{\bf Appendix \thesubappendixc. {\kern1pt \bfit #1}}
	\par\vspace{5pt}}
\newcommand{\subsubappendix}[1] {\vspace{12pt}
        \refstepcounter{subsubappendixc}
        \noindent{\rm Appendix \thesubsubappendixc. {\kern1pt \tenit #1}}
	\par\vspace{5pt}}

\newcommand{\textlineskip}{\baselineskip=13pt}
\newcommand{\smalllineskip}{\baselineskip=10pt}

\def\eightcirc{
\begin{picture}(0,0)
\put(4.4,1.8){\circle{6.5}}
\end{picture}}
\def\eightcopyright{\eightcirc\kern2.7pt\hbox{\eightrm c}} 

\newcommand{\copyrightheading}[1]
	{\vspace*{-2.5cm}\smalllineskip{\flushleft
	{\footnotesize International Journal of Modern Physics A, #1}\\
	{\footnotesize $\eightcopyright$\, World Scientific Publishing
	 Company}\\
	 }}

\newcommand{\publisher}[2]{{\begin{center}\footnotesize\smalllineskip 
	Received #1\\
	Revised #2
	\end{center}
	}}

\def\abstracts#1#2#3{{
	\centering{\begin{minipage}{4.5in}\baselineskip=10pt\footnotesize
	\parindent=0pt #1\par 
	\parindent=15pt #2\par
	\parindent=15pt #3
	\end{minipage}}\par}} 

\newcommand{\bibit}{\nineit}

\renewenvironment{thebibliography}[1]
	{\frenchspacing
	 \ninerm\baselineskip=11pt
	 \begin{list}{\arabic{enumi}.}
	{\usecounter{enumi}\setlength{\parsep}{0pt}
	 \setlength{\leftmargin 12.7pt}{\rightmargin 0pt} %FOR 1--9 ITEMS
	 \setlength{\itemsep}{0pt} \settowidth
	{\labelwidth}{#1.}\sloppy}}{\end{list}}

\newcounter{itemlistc}
\newcounter{romanlistc}
\newcounter{alphlistc}
\newcounter{arabiclistc}

\newcommand{\fcaption}[1]{
        \refstepcounter{figure}
        \setbox\@tempboxa = \hbox{\footnotesize Fig.~\thefigure. #1}
        \ifdim \wd\@tempboxa > 5in
           {\begin{center}
        \parbox{5in}{\footnotesize\smalllineskip Fig.~\thefigure. #1}
            \end{center}}
        \else
             {\begin{center}
             {\footnotesize Fig.~\thefigure. #1}
              \end{center}}
        \fi}

\newcommand{\tcaption}[1]{
        \refstepcounter{table}
        \setbox\@tempboxa = \hbox{\footnotesize Table~\thetable. #1}
        \ifdim \wd\@tempboxa > 5in
           {\begin{center}
        \parbox{5in}{\footnotesize\smalllineskip Table~\thetable. #1}
            \end{center}}
        \else
             {\begin{center}
             {\footnotesize Table~\thetable. #1}
              \end{center}}
        \fi}

\def\@citex[#1]#2{\if@filesw\immediate\write\@auxout
	{\string\citation{#2}}\fi
\def\@citea{}\@cite{\@for\@citeb:=#2\do
	{\@citea\def\@citea{,}\@ifundefined
	{b@\@citeb}{{\bf ?}\@warning
	{Citation `\@citeb' on page \thepage \space undefined}}
	{\csname b@\@citeb\endcsname}}}{#1}}

\newif\if@cghi
\def\cite{\@cghitrue\@ifnextchar [{\@tempswatrue
	\@citex}{\@tempswafalse\@citex[]}}
\def\citelow{\@cghifalse\@ifnextchar [{\@tempswatrue
	\@citex}{\@tempswafalse\@citex[]}}
\def\@cite#1#2{{$\null^{#1}$\if@tempswa\typeout
	{IJCGA warning: optional citation argument 
	ignored: `#2'} \fi}}

\def\pmb#1{\setbox0=\hbox{#1}
	\kern-.025em\copy0\kern-\wd0
	\kern.05em\copy0\kern-\wd0
	\kern-.025em\raise.0433em\box0}

\def\fnt#1#2{\footnotetext{\kern-.3em
	{$^{\mbox{\scriptsize #1}}$}{#2}}}

\def\fpage#1{\begingroup
\voffset=.3in
\thispagestyle{empty}\begin{table}[b]\centerline{\footnotesize #1}
	\end{table}\endgroup}

\font\tenrm=cmr10
\font\tenit=cmti10 
\font\tenbf=cmbx10
\font\bfit=cmbxti10 at 10pt
\font\ninerm=cmr9
\font\nineit=cmti9

\font\eightrm=cmr8

\def\qed{\hbox{${\vcenter{\vbox{			%HOLLOW SQUARE
   \hrule height 0.4pt\hbox{\vrule width 0.4pt height 6pt
   \kern5pt\vrule width 0.4pt}\hrule height 0.4pt}}}$}}

\renewcommand{\thefootnote}{\fnsymbol{footnote}}	%USE SYMBOLIC FOOTNOTE

\begin{document}

%\runninghead{Instructions for Typesetting Camera-Ready
%Manuscripts $\ldots$} {Instructions for Typesetting Camera-Ready
%Manuscripts $\ldots$}

\normalsize\textlineskip
\thispagestyle{empty}
\setcounter{page}{1}

\copyrightheading{}			%{Vol. 0, No. 0 (1993) 000--000}

\vspace*{0.88truein}

\fpage{1}
\centerline{\bf SPECTROSCOPY AT B-FACTORIES USING HARD} 
\vspace*{0.035truein}
\centerline{\bf PHOTON EMISSION} 
\vspace*{0.37truein}
\centerline{\footnotesize M.~BENAYOUN$^{a}$, S.~I.~EIDELMAN$^{a,b}$,
V.~N.~IVANCHENKO$^{a,b}$, Z.~K.~SILAGADZE$^{b}$ }
\vspace*{0.015truein}
\centerline{$^{a}$LPNHE des Universit\'es Paris VI et VII--IN2P3, Paris,
         France\\}
\vspace*{0.015truein}
\centerline{$^{b}$Budker Institute of Nuclear Physics, 630090 Novosibirsk, 
         Russia \\}
\baselineskip=10pt
\vspace*{0.225truein}
\publisher{(received date)}{(revised date)}

\vspace*{0.21truein}
\abstracts{
The process of hard photon  emission by initial
electrons (positrons)
at B-factories 
is discussed. It is shown that studies of the bottomonium spectroscopy
will be feasible for the  planned integrated luminosity
of the B-factory experiments.}
{}{}

\textheight=7.8truein
\setcounter{footnote}{0}
\renewcommand{\thefootnote}{\alph{footnote}}
\vspace*{1pt}\textlineskip	%) USE THIS MEASUREMENT WHEN THERE IS
\section{Introduction}		%) A SECTION HEADING
\vspace*{-0.5pt}
\noindent
Heavy quark physics is one of the frontier areas in 
studies of the fundamental properties of matter. That is why a new 
generation of $e^+e^-$ colliders (B-factories) has been designed.
The CLEO-III \cite{CLEO3}, BELLE \cite{BELLE}, and 
BABAR \cite{BABAR} experiments
are running  since 1999. Their main goal 
is the  precise  investigation of the Cabibbo-Kobayashi-Maskawa
matrix and, first of all,  CP-violation in the  b-quark sector.
To achieve this goal, most of the time 
will be spent on 
running at the energy of the $\Upsilon (4S)$ resonance.
At the same time there are 
physical tasks  \cite{VZ} related with the  spectroscopy of
quarkonium systems which require  scanning   the energy region of
 $\Upsilon (1S)$, $\Upsilon (2S)$, $\Upsilon (3S)$.
Such
experiments  were  performed 
at DORIS, CESR, and VEPP-4 
\cite{PDG}. The main part of the bottomonium family 
has been observed, their masses and
main decay modes have been measured. The list of  experiments with 
the largest integrated luminosity is shown in Table \ref{T0}.
\begin{table}[htbp]
\tcaption{
Selected experiments 
with the largest integrated luminosity 
in which transitions between different
$b\overline{b}$ states have been studied.
The integrated luminosity (L) and 
corresponding  number of 
produced initial $\Upsilon (2S)$ or $\Upsilon (3S)$ mesons (N)
are shown.
}
\centerline{\footnotesize\smalllineskip
\begin{tabular}{l c c c}\\
\hline 
\hline 
Process & Detector & L, $pb^{-1}$ & $N, 10^{6}$ \\ 
\hline
$\Upsilon(3S)\to\chi_b\gamma$ & CLEO-II \cite{CLEO91} & 116 & 0.40  \\
\hline
$\Upsilon(3S)\to\chi_b\gamma$ & CUSB-II \cite{CUSB2} & 288 & 0.99  \\
$\Upsilon(3S)\to\Upsilon(2S)\pi^0\pi^0$ & & & \\
$\Upsilon(3S)\to\Upsilon(1S)\pi^0\pi^0$ & & & \\
\hline
$\Upsilon(3S)\to\Upsilon(2S)\pi^+\pi^-$ & CLEO-II \cite{CLEO94} & 130 & 0.47\\
$\Upsilon(3S)\to\Upsilon(2S)\pi^0\pi^0$ & & & \\
$\Upsilon(3S)\to\Upsilon(1S)\pi^+\pi^-$ & & & \\
$\Upsilon(3S)\to\Upsilon(1S)\pi^0\pi^0$ & & & \\
\hline
$\Upsilon(2S)\to\Upsilon(1S)\pi^+\pi^-$& CLEO-II \cite{CLEO2} & 79 & 0.49\\
$\Upsilon(2S)\to\Upsilon(1S)\pi^0\pi^0$ & & & \\
\hline 
\hline\\
\end{tabular}}
\label{T0}
\end{table}
However, the accuracy of the measured parameters of the
bottomonium family
is not very high, preventing from the  precise comparison
between the data and 
different model predictions for the mass spectrum,  decay rates,
and dynamics of transitions. Moreover, some 
predicted states ($\eta_b(2S)$, $\eta_b(1S)$, $h_b(1P)$) have not yet been 
observed or are not well established.
Thus, new experiments 
are necessary which will provide complete information about
all states below the open flavor threshold.

Unfortunately,
scans of the $\Upsilon (1S)$, $\Upsilon (2S)$, $\Upsilon (3S)$
are not foreseen at
asymmetric B-factory  
experiments. Of course, it can be done in a traditional way by
CLEO-III but that will decrease the total integrated luminosity 
which is expected to be collected
by CLEO-III  at $\Upsilon (4S)$. That is why we consider a possibility 
to perform $\Upsilon$-spectroscopy studies at 
the $\Upsilon (4S)$ energy
using the emission of a hard photon  by the electron or
the positron. This process 
 is well known, it
 changes the value of the measured cross section 
(so called ``radiative corrections'')
and
has to be taken into account in practically  all $e^+e^-$ experiments.

Different possibilities 
of using radiative photons are now
under intensive discussions.
It was suggested to measure the longitudinal structure
function $F_L(x,Q^2)$ at HERA using
events with the emission of a
hard photon collinear to the
incident lepton beam
\cite{CRASNY,FAVART}. Such a measurement has recently been performed
by H1 \cite{H1}.
Other authors  \cite{ARBUZ,ARBUZ1,BKM,KM,SPAG} 
suggest to use a similar process for precise 
determination of R at the  DA$\Phi$NE $\phi$-factory. 
In the present work 
the following reaction has been  studied:  
\begin{equation}
{e^+e^- \to\gamma V\to\gamma f,} 
\label{REAC}
\end{equation}
where V is one of the vector  resonances
$\Upsilon(3S)$, $\Upsilon(2S)$,
$\Upsilon(1S)$ etc. decaying to a final state $f$.
Our main task is to show the practical feasibility of using
such  processes for spectroscopy studies
at B-factories in near future.

\section{Calculation of the Production Cross Sections}
\noindent
At first order of quantum electrodynamics (QED) the process 
(\ref{REAC}) is described by two diagrams (Fig.1). 
The Born term for this process 
can be obtained using 
the quasi-real electron method \cite{BFKh,BM}
\begin{equation}
\frac{d\sigma(s,x)}{dx~d\cos{\theta}} = 
\frac {2\alpha}{\pi x}\cdot \frac{(1-x+\frac{x^2}{2}) \sin^2{\theta}}
{(\sin^2{\theta} + \frac{m_e^2}{E^2}\cos^2{\theta})^2} 
\cdot \sigma_0(s(1-x)),
\label{SIG0}
\end{equation}
where $s = 4E^2$, E is the beam energy  in the 
center of mass system of the electron and positron,
$m_e$ is the electron mass, 
$\alpha$ is
the fine structure constant,
$x=E_{\gamma}/E$ is the fraction of 
the beam energy taken by the radiative photon with the energy $E_{\gamma}$,
$\theta$  is the photon emission
angle with respect to the beam ($0<\theta< \pi $),
$\sigma_0(s)$ is the cross section of hadronic production in $e^+e^-$
annihilation.

\begin{figure}[htbp]
\vspace*{13pt}
%\centerline{\vbox{\hrule width 5cm height0.001pt}}
%\vspace*{1.4truein}		%ORIGINAL SIZE=1.6TRUEIN x 100% - 0.2TRUEIN
%\input {afig.tex}
%\centerline{\vbox{\hrule width 5cm height0.001pt}}
  \epsfig{figure=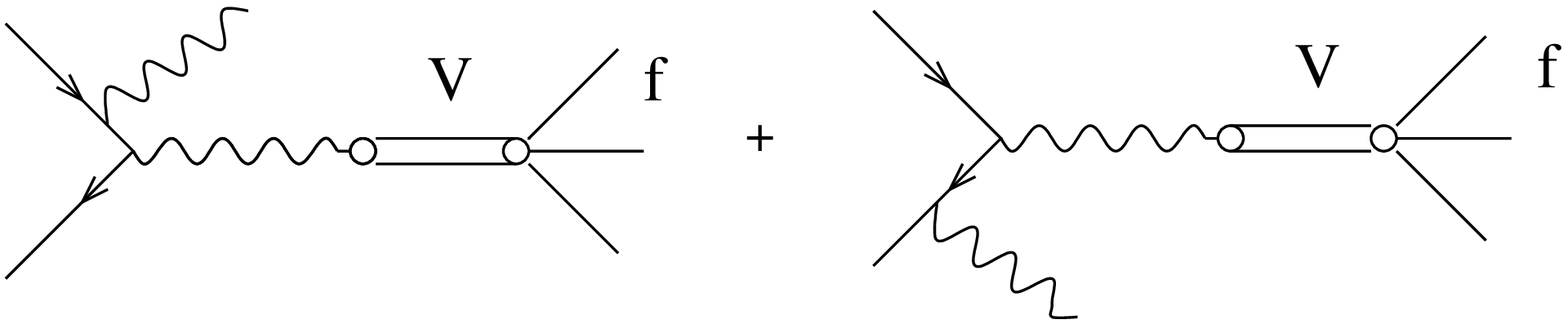,height=3cm,width=12cm}
\vspace*{13pt}
\fcaption{Main diagrams describing the process of hard photon emission.}
\end{figure}

If one performs  integration  over $\theta$ in (\ref{SIG0}), the following 
energy dependence is obtained:
\begin{equation}
{\frac{d\sigma(s,x)}{dx} = 
W(s,x)\cdot \sigma_0(s(1-x)),}
\label{SIG}
\end{equation}
where  $W(s,x)$ is 
the probability function of the photon emission which
 can be written:
\begin{equation}
{W(s,x)=\frac{2\alpha}{\pi\cdot x}
\cdot (L-1)\cdot (1-x+\frac{x^2}{2}),
\; \; L=2\ln{\frac{\sqrt{s}}{m_e}}.}
\label{ENER}
\end{equation}

The Born
cross section of the narrow 
vector resonance $V$ production is given by the 
standard Breit-Wigner formula
\begin{equation}
{\sigma_0(s)=\frac{12\pi B_{ee}}{m^2_V}\cdot\frac{m^2_V\Gamma^2_V}
{(s-m^2_V)^2+m^2_V\Gamma^2_V},}
\label{SIGBW}
\end{equation}
where $m_V$ and $\Gamma_V$ are the resonance mass and  width respectively,
$B_{ee}$ is the branching fraction of the $V\to e^+e^-$ decay. 
If the resonance is narrow, 
the photon 
energy spectrum 
in reaction (\ref{REAC}) is also narrow. The central value
of the photon energy fraction is $x_V = (1-m^2_V/s)$. 
For a very narrow resonance, one can replace
\begin{equation}
{\frac{m_V\Gamma_V}{(s-m_V^2)^2+m_V^2\Gamma_V^2} \longrightarrow
\pi \delta(s-m_V^2)}
\label{BW0}
\end{equation}
\noindent
The total cross section $\sigma_V(s)$ of the process (\ref{REAC}) 
 can be found by the integration of
(\ref{SIG}) in the region around $x_V$ using (\ref{SIGBW}) and (\ref{BW0})
\begin{equation}
{\sigma_V(s)=\frac{12\pi^2 B_{ee}\Gamma_V}{m_V \cdot s}
\cdot W(s,x_V).}
\label{SIGV}
\end{equation}

Assuming that the experiment is carried out at the $\Upsilon(4S)$
energy, the 
production cross sections of vector mesons with quantum 
numbers $J^{PC}=1^{--}$ can be calculated
using (\ref{SIGV}) and PDG values for the meson
masses and widths \cite{PDG}. 
The results are shown in 
Table \ref{T1}. In addition, we present in this Table
the numbers of produced events which
correspond to a collected integrated luminosity of $10~fb^{-1}$.

\begin{table}[htbp]
\tcaption{
Cross sections for radiative production of
$J^{PC}=1^{--}$ mesons at the $\Upsilon(4S)$ energy: $\sigma_V^I$ 
is the cross section 
calculated at first order; $\sigma_V^{II}$ is the cross section
calculated with leading second order
corrections; $N_{total}$ is 
the total number of mesons produced
in the experiment with the integrated luminosity of  $10~fb^{-1}$; 
$N_{\gamma l^+l^-}$ is
the number of lepton decays for radiative production. The estimation of the  
$\Upsilon(4S)$ production cross section 
and the number of produced radiative Bhabha events and $\tau$-lepton 
pairs 
at $\Upsilon(4S)$
are shown for  comparison. 
}
\centerline{\footnotesize\smalllineskip
\begin{tabular}{l c c c c c}
\hline 
\hline 
Meson & $\sigma_V^I$~(nb) &$\sigma_V^{II}$~(nb) & $N_{total}, 10^{6}$ & 
$ N_{\gamma e^+e^-}, 10^{3} $ & 
$ N_{\gamma\tau^+\tau^-}, 10^{3}$ \\
\hline
$\Upsilon(4S)$ &   - & 3.40 & 34 & $4\cdot10^6$ & $1.1\cdot10^4$  \\
$\Upsilon(3S)$ &   0.038 &0.031 & 0.31 & 5.59 & 5.55  \\
$\Upsilon(2S)$ &   0.016 &0.015 & 0.15 & 1.83 & 1.81 \\
$\Upsilon(1S)$ &   0.021 &0.019 & 0.19 & 4.71 & 4.67 \\
$\psi(2S)    $ &   0.013 &0.014 & 0.14 & 1.22 & 0.47 \\
$J/\psi(1S)  $ &   0.034 &0.036 & 0.36 & 21.5 & 0 \\
$\phi$         &   0.024 &0.027 & 0.27 & 0.08 & 0 \\
$\omega$       &   0.014 &0.016 & 0.16 & 0.01 & 0 \\
$\rho$         &   0.160 &0.182 & 1.82 & 0.08 & 0 \\
\hline 
\hline\\
\end{tabular}}
\label{T1}
\end{table}

These results  demonstrate that radiative production 
cross sections have the same order of magnitude for all quarkonium states. 
These values show the practical feasibility of using
such processes for spectroscopy studies
at B-factories in near future. 
At the same time we have to point out 
that  reaction (\ref{REAC})
could provide some background for various
processes which will be studied at B-factory experiments, 
particularly for two photon processes and $\tau$-lepton decays.
That is why such processes have to be investigated 
at asymmetric B-factories,
even if CLEO-III will perform energy scan of 
$\Upsilon (2S)$ or  $\Upsilon (3S)$ regions  
with a high integrated luminosity.

The expression for the differential cross sections (\ref{SIG0})
allows to estimate
that about 50~\% of photons are emitted within the angle of 20~mrad
to the beam axis. 
Using the parameters of B-factories \cite{PDG}, one can 
calculate the values of velocity in radiative production (Table \ref{T2})
for reaction (\ref{REAC}) in the case
when photons are directed along the beam.
Kinematic conditions at 
 B-factories are slightly different, so produced vector 
mesons will have different boosts. Note that $\Upsilon$-mesons at 
asymmetric B-factories will fly in the direction of 
the high energy beam. Light
mesons will have both directions.

\begin{table}[htbp]
\tcaption{Velocity $\beta=v/c$ of the meson in
the radiative production at 
 B-factories. For the asymmetric factories
two values are shown corresponding to the  
 radiative photon 
emitted by the positron or by the electron.
Photon direction coincides with the beam. The values of the
$\Upsilon(4S)$ velocity are shown for comparison.
}
\centerline{\footnotesize\smalllineskip
\begin{tabular}{l l l l}
\hline 
\hline 
Factory & CESR & KEKB & PEP-II \\
\hline 
$\Upsilon(4S)$ &   0 & 0.391 & 0.488  \\
$\Upsilon(3S)$ &   0.023 & 0.410, 0.373 & 0.503, 0.472  \\
$\Upsilon(2S)$ &   0.056 & 0.436, 0.344 & 0.527, 0.447  \\
$\Upsilon(1S)$ &   0.113 & 0.482, 0.292 & 0.567, 0.399  \\
$\psi(2S)    $ &   0.784 & 0.899, 0.566 & 0.920, 0.478  \\
$J/\psi(1S)  $ &   0.843 & 0.928, 0.673 & 0.942, 0.601  \\
$\phi$         &   0.982 & 0.992, 0.958 & 0.994, 0.947  \\
$\omega$       &   0.989 & 0.995, 0.975 & 0.996, 0.969  \\
$\rho$         &   0.990 & 0.995, 0.976 & 0.996, 0.970  \\
\hline 
\hline\\
\end{tabular}}
\label{T2}
\end{table}

\newpage

\section{Remarks on the Leading $\alpha^2$ Corrections}
\noindent
In the previous section the cross section at first order in $\alpha$ 
has been used  to estimate the
main effects of hard photon emission. 
Expressions which take into account the leading $\alpha^2$ contributions 
are also well known \cite{YFS,JS,JW,FADKUR,NT,CKT}. 
Equation (\ref{SIG}) gives a correct estimate for the radiative photon 
emission cross section with about 10-20~\% precision. Two important 
improvements are necessary in this formula if better accuracy
is required.
First of all, the cross section as given by Eq. (\ref{SIG}) is
infrared divergent. To cure this shortcoming, it is necessary to sum all
diagrams with soft multi-photon emission. It is well known \cite{YFS} that
this procedure leads to the soft-photon exponent. A  practically convenient
(but not completely correct) recipee for this exponentiation in a narrow 
resonance case was given in \cite{JS}. Exponentiation of soft photons in
the case of Bonneau and Martin cross section was further considered in
\cite{JW} with an explicit Monte Carlo algorithm for this exponentiation.

Some second order corrections containing large logarithms can also 
contribute up to a several percent level. The structure function method 
suggested in \cite{FADKUR} is a popular tool to calculate finite-order
leading logarithmic corrections. This formalism also gives 
the factorized form (\ref{SIG}) for the radiative photon emission
cross section. $W(s,x)$ function, the so called QED ``radiator'', consists
of two parts. The exponentiated part accounts for soft multi-photon 
emission, while the remaining one takes into account hard collinear 
bremsstrahlung
in the leading logarithmic approximation. Up to order $\alpha^2$, the
radiator looks like \cite{NT,BERENDS}
\begin{eqnarray}
W(s,x) & = &
\Delta \cdot \beta x^{\beta -1}-\frac{\beta}{2} (2-x) + \nonumber \\ 
& & \frac{\beta^2}{8}
\left \{ (2-x)[3\ln{(1-x)}-4\ln{x}]-4\frac{\ln{(1-x)}}{x}-6+x \right \},
\nonumber \\
\Delta & = &
1+\frac{\alpha}{\pi} \left ( \frac{3}{2}L+\frac{1}{3}\pi^2-2 \right ) +
\left ( \frac{\alpha}{\pi}\right )^2 \delta_2, \nonumber \\
\delta_2 & = & \left (\frac{9}{8}-2\zeta_2\right ) L^2 -   
\left ( \frac{45}{16}-\frac{11}{2}\zeta_2-3\zeta_3\right ) L- 
\nonumber \\ & & 
\frac{6}{5}\zeta_2^2-\frac{9}{2}\zeta_3-6\zeta_2 \ln{2}+
\frac{3}{8}\zeta_2 + \frac{57}{12}, \nonumber \\
\beta & = & \frac{2\alpha}{\pi}(L-1), 
\; \; \zeta_2=1.64493407,\; \; \zeta_3=1.2020569.
\label{AL4}
\end{eqnarray}
\noindent 
Note that some  $\alpha^3$ contributions in the QED radiator 
are also known \cite{MNP}, but they are irrelevant for our goals.
Using expression (\ref{AL4}) and formula (\ref{SIG}), one can obtain  
the differential cross section of
radiative production at second order in $\alpha$. 
 The  corresponding total cross section is given 
by the relation (\ref{SIGV}).
The results  of calculations
taking into account leading $\alpha^2$ corrections are shown
in Table \ref{T1}. The difference with the first order estimation is  20~\%
for $\Upsilon(3S)$ production and (5-10)~\% for the 
production of other resonances.

The angular distribution of the emitted photon given by Eq.
(\ref{SIG0}) is valid  for soft enough photons and small emission
angles only. A more general result 
which can be obtained by the method of 
Bonneau and Martin \cite{BM} up to $m_e^2/s$ terms, is
\begin{eqnarray}
\frac{d\sigma(s,x)}{dx~d\cos{\theta}} & = & 
\frac{2\alpha}{\pi x}\cdot (1-x+\frac{x^2}{2})\cdot \sigma_0(s(1-x))
\cdot P(\theta), 
\nonumber \\ 
P(\theta) & = &
\frac 
%\left \{
{\sin^2{\theta}-\frac{x^2\sin^4{\theta}}{2(x^2-2x+2)}-
\frac{m_e^2}{E^2}~\frac{(1-2x)\sin^2{\theta}-x^2\cos^4{\theta}}{x^2-2x+2}}
%\right \}
{\left ( \sin^2{\theta}+\frac{m_e^2}{E^2}\cos^2{\theta}
\right )^2 }.
\label{phad} 
\end{eqnarray}
\noindent 
This result is in agreement with one of the first works
 \cite{BAIER} devoted to the radiative process
$e^+e^-\to \mu^+\mu^-\gamma$. The $m_e^2/s$ terms 
in  (\ref{phad}) become important for very small angles $\theta$ only.
Nevertheless, since these terms are drastically peaked, they give 
non-negligible contribution to $d\sigma /dx$ and their presence 
is necessary to obtain the correct form at order $\alpha$ for
this differential cross section, as given by equations 
(\ref{SIG}) and (\ref{ENER}).
Note that expressions presented in Refs.\cite{BKM,KM,MEX} do not
include such terms.

From (\ref{phad}) one can obtain the probability for the hard
photon to be emitted inside the cone of opening the angle $\theta_m$ 
\begin{equation}
{\cal P}(0\le \theta \le \theta_m)=\frac{h(\theta_m)}{h(\pi)}, \; \; \;
h(\theta_m)=\int\limits_0^{\theta_m} P(\theta)\sin{\theta}
d\theta\; ,
\label{PRB} 
\end{equation}
\noindent where
\begin{eqnarray}
h(\theta) & = &
\frac{L-1}{2}+\frac{m_e^2}{2E^2}~
\frac{\cos{\theta}}{\sin^2{\theta} + \frac{m_e^2}{E^2}\cos^2{\theta}}
-\frac{1}{2}\ln{\frac{1+\sqrt{1-\frac{m_e^2}{E^2}}\cos{\theta}}
{1-\sqrt{1-\frac{m_e^2}{E^2}}\cos{\theta}}}+ \nonumber \\
& & \frac{x^2\cos{\theta}}{2(x^2-2x+2)} \left ( 1-\frac{m_e^2}{E^2}
\frac{1}{\sin^2{\theta} + \frac{m_e^2}{E^2}\cos^2{\theta}} \right ) .
\label{ANG} 
\end{eqnarray}
\noindent Note that $h(\pi)=L-1$.

\section{Prospects for Spectroscopy of Bottomonium}
\noindent
To study the bottomonium spectroscopy, two types of events  
can be considered for which background
will be small. The first one can be referred to as ``tagged photon'' events.
In these events a hard photon is emitted at a  
large angle with respect to the 
beam axis and is 
recorded by the main calorimeter 
of the detector. 
Using the high energy and position
resolution of the calorimeter for photons, one can
reconstruct the recoil energy of the hadronic system in 
the  reaction (\ref{REAC}) and thereby precisely identify 
the produced vector meson $V$. In this case the investigation
of its decay modes becomes possible without requiring that all final
particles from the $V$ decay be detected.
If another hard collinear photon is additionally emitted, the recoil
energy spectrum is smeared and acquires some tails, but this effect should 
not be large.

Formulae (\ref{PRB}) and (\ref{ANG})  allow  to estimate the 
``tagged photon'' detection efficiency for the processes (\ref{REAC}).
To do that, the probability of the hard photon emission 
at angle $\theta$  relative
to the beam direction is calculated
($\theta_{min} < \theta < \theta_{max}$).
The energy of the photon is fixed by the mass of the 
produced resonance $x=x_V$. 
The maximum and minimum angles
$\theta_{max}$, $\theta_{min}$ are defined by the geometry of the  
electromagnetic calorimeter of the specific detector
\cite{CLEO3,BELLE,BABAR}. Results of the calculation  
are shown in Table \ref{TT}.

\begin{table}[htbp]
\tcaption{The detection 
efficiency for the hard photon in 
calorimeters of the B-factory detectors.
}
\centerline{\footnotesize\smalllineskip
\begin{tabular}{l l l l c}
\hline
\hline 
Detector & $\theta_{min}, degrees$ & $\theta_{max}, degrees$ & Meson &
 $\epsilon, \%$  \\
\hline 
 & &  &  $\Upsilon(3S)$ & 13.9 \\
 & &  &  $\Upsilon(2S)$ & 13.9\\
 & &  &  $\Upsilon(1S)$ & 13.8\\
 & &  &  $\Psi(2S)$ & 10.5\\
CLEO-III &  30.0 & 150.0  & $J/\Psi (1S)$ & 10.1 \\
 & &  & $\phi$   & 9.5\\
 & &  & $\omega$ & 9.4 \\
 & &  & $\rho$   & 9.4\\
\hline
 & &  &  $\Upsilon(3S)$ & 20.0 \\
 & &  &  $\Upsilon(2S)$ & 19.9 \\
 & &  &  $\Upsilon(1S)$ & 19.8\\
 & &  &  $\Psi(2S)$ & 16.1 \\
BELLE &   18.3   & 163.7 & $J/\Psi (1S)$ & 15.8  \\
 & &  & $\phi$   & 15.0\\
 & &  & $\omega$ & 15.0\\
 & &  & $\rho$   & 15.0\\
\hline
 & &  &  $\Upsilon(3S)$ & 15.9\\
 & &  &  $\Upsilon(2S)$ & 15.9\\
 & &  &  $\Upsilon(1S)$ & 15.8\\
 & &  &  $\Psi(2S)$ & 12.3\\
BABAR &   26.5   & 156.3  & $J/\Psi (1S)$ & 12.0 \\
 & &  & $\phi$   & 11.2\\
 & &  & $\omega$ & 11.2\\
 & &  & $\rho$   & 11.2\\
\hline 
\hline\\
\end{tabular}}
\label{TT}
\end{table}

As mentioned above,
the main part of hard photons is emitted along
the  beam axis. Thus, this photon will not
fire the detector calorimeter. In this case complete reconstruction of
the hadronic system from the reaction (\ref{REAC}) is necessary.
It was recently shown by CLEO-II  \cite{CLEO2} that, 
studying events with lepton  and pion pairs in the final state, one
can obtain a practically clean sample of pion transitions between 
$\Upsilon(2S)$ and $\Upsilon(1S)$ resonances. The estimated
number of events of  such type in the reaction (\ref{REAC}) at $\Upsilon(4S)$
is shown in Table \ref{T3}.

\begin{table}[htbp]
\tcaption{The estimated number of events of hadron transitions
with two pions and a lepton pair in the final state 
for an experiment with the integrated luminosity of $10~fb^{-1}$. 
}
\centerline{\footnotesize\smalllineskip
\begin{tabular}{l c}
\hline 
\hline 
Reaction & N events \\
\hline 
$\Upsilon(3S)\to\Upsilon(2S)\pi\pi\to l^+l^-\pi\pi$ & 428  \\
$\Upsilon(3S)\to\Upsilon(1S)\pi\pi\to l^+l^-\pi\pi$ & 1248  \\
$\Upsilon(2S)\to\Upsilon(1S)\pi\pi\to l^+l^-\pi\pi$ & 2268  \\
\hline 
\hline\\
\end{tabular}}
\label{T3}
\end{table}

One of the important tasks for the bottomonium spectroscopy 
is to study radiative transitions, when $\Upsilon(3S)$ or $\Upsilon(2S)$
decays to one of the $\chi_b$ states and after another radiative
decay $\Upsilon(2S)$ or $\Upsilon(1S)$ is produced.
If the final $\Upsilon(2S)$ or $\Upsilon(1S)$ decays into a
lepton pair, these reactions possess a clear signature 
``two photons and lepton pair''
allowing obvious background rejection. The estimated number of 
produced events with  a gamma 
transition is shown in Table \ref{T4}.

\begin{table}[htbp]
\tcaption{The estimated number  of radiative decay events
with two photons and a lepton pair in the final state
for experiment with the integrated luminosity of $10~fb^{-1}$. 
}
\centerline{\footnotesize\smalllineskip
\begin{tabular}{l c}
\hline 
\hline 
Reaction & N events \\
\hline 
$ \Upsilon(3S)\to\chi_{b2}(2P)\gamma\to\Upsilon(2S) \gamma\gamma $ & 164  \\
$ \Upsilon(3S)\to\chi_{b1}(2P)\gamma\to\Upsilon(2S) \gamma\gamma $ & 212  \\
$ \Upsilon(3S)\to\chi_{b0}(2P)\gamma\to\Upsilon(2S) \gamma\gamma $ &  22  \\
$ \Upsilon(3S)\to\chi_{b2}(2P)\gamma\to\Upsilon(1S) \gamma\gamma $ & 183  \\
$ \Upsilon(3S)\to\chi_{b1}(2P)\gamma\to\Upsilon(1S) \gamma\gamma $ & 22  \\
$ \Upsilon(3S)\to\chi_{b0}(2P)\gamma\to\Upsilon(1S) \gamma\gamma $ &  9  \\
$ \Upsilon(2S)\to\chi_{b2}(1P)\gamma\to\Upsilon(1S) \gamma\gamma $ & 122  \\
$ \Upsilon(2S)\to\chi_{b1}(1P)\gamma\to\Upsilon(1S) \gamma\gamma $ & 192  \\
$ \Upsilon(2S)\to\chi_{b0}(1P)\gamma\to\Upsilon(1S) \gamma\gamma $ & 10  \\
\hline 
\hline\\
\end{tabular}}
\label{T4}
\end{table}

The number of estimated events with 
$\Upsilon (1S), \Upsilon (2S), \Upsilon (3S)$ decays
shown in Table \ref{T1} is smaller than the one 
collected by a standard method of scanning each resonance
in CLEO-II experiments \cite{CLEO91,CLEO94,CLEO2}.
But the integrated 
luminosity to be collected by the designed B-factories
is at least one order of magnitude higher than that  
in our estimations. This means that 
reaction (\ref{REAC}) can serve as a real source of all bottomonium states
and provide new independent information
complementary  to the existing data 
as well as to that expected from future experiments 
with CLEO-III \cite{CLEO3}.

\section{Conclusion}
\noindent
Estimations performed in this work have demonstrated feasibility 
of using  a radiative photon for studies of the bottomonium spectroscopy at
B-factories. These processes have to be taken into account 
as a possible background for various reactions which will
be studied in these experiments at the  $\Upsilon(4S)$ energy.
If the integrated luminosity
of about $100~fb^{-1}$ is collected, the number of events for
spectroscopy studies with clean signature will be higher than that  
collected now by the traditional method of scanning the 
resonance energy range. 

The same method can be considered to study  
$\rho$ and $\omega$ meson decays at the DA$\Phi$NE $\phi$-factory. 
For example, for an integrated luminosity
of about $300~pb^{-1}$, the number of
$\omega\gamma$ events is $10^6$ and that of $\rho\gamma$ is $10^7$.

\section{Acknowledgments}
\noindent
SIE was supported by
the Division des Affaires Internationales of IN2P3 and would like to
thank
the LPNHE Laboratory for its hospitality; VNI was supported by
the Direction des Affaires Internationales of CNRS.
Both SIE and VNI are grateful to
Eliane Perret (IN2P3) and
Marcel Banner (LPNHE) for their help and support.
We are grateful to Kirill Melnikov and Alexei Vasiljev for
stimulating discussions.

\nonumsection{References}

\end{document}